\input tables
% determine hypertex mode
\newif\iflanl
\openin 1 lanlmac
\ifeof 1 \lanlfalse \else \lanltrue \fi
\closein 1
\iflanl
    \input lanlmac
\else
    \message{[lanlmac not found - use harvmac instead}
    \input harvmac
    \fi
\newif\ifhypertex
\ifx\hyperdef\UnDeFiNeD
    \hypertexfalse
    \message{[HYPERTEX MODE OFF}
    \def\hyperref#1#2#3#4{#4}
    \def\hyperdef#1#2#3#4{#4}
    \def\hypernoname{}
    \def\e@tf@ur#1{}
    \def\eprt#1{{\tt #1}}
    \def\CERN{\centerline{CERN, CH--1211 Geneva 23, Switzerland}}
    \def\wl{W.\ Lerche}
\else
    \hypertextrue
    \message{[HYPERTEX MODE ON}
%hypertex links to xxx.lanl.gov:
%  \def\hth/#1#2#3#4#5#6#7{\special{html:<a
%   href="http://xxx.lanl.gov/abs/hep-th/#1#2#3#4#5#6#7">}
%  {\tt hep-th/#1#2#3#4#5#6#7}\special{html:</a>}}
\def\eprt#1{{\tt
#1}}
\def\CERN{\centerline{

Theory Division, CERN, Geneva, Switzerland}}
\def\wl{
 W.\ Lerche}
\fi
%%%%%%%%%%%%%%%%%%%%%%% %%%%%%%%%%%%%%%%%%%%%%%
\newif\ifdraft

\noblackbox
\catcode`\@=11
\newif\iffrontpage
%%%%%%%%%%%%%%%%%%% %%%%%%%%%%%%%%%%%%%%%%%%%%%%%%%%%%%%%%%%%%%%%%
%%%%% sizes, offsets etc
%%%%%%%%%%%%%%%%%%% %%%%%%%%%%%%%%%%%%%%%%%%%%%%%%%%%%%%%%%%%%%%%%
\ifx\answ\bigans
\def\titleft{\titla}
\magnification=1200\baselineskip=14pt plus 2pt minus 1pt
%
%%%%% unreduced mode: %%%%
%\voffset=0.35truein\hoffset=0.250truein
\advance\hoffset by-0.075truein
\advance\voffset by1.truecm
\hsize=6.15truein\vsize=600.truept\hsbody=\hsize\hstitle=\hsize
\else\let\lr=L
\def\titleft{\titla}
\magnification=1000\baselineskip=14pt plus 2pt minus 1pt
%
%%%%% reduced mode: %%%%%%%
\hoffset=-0.75truein\voffset=-.0truein
%?\hoffset=-.25truein\voffset=-.0truein
\vsize=6.5truein
\hstitle=8.truein\hsbody=4.75truein
\fullhsize=10truein\hsize=\hsbody
\fi
\parskip=4pt plus 15pt minus 1pt
%
%%%%%%%%%%%%%%%%%%% %%%%%%%%%%%%%%%%%%%%%%%%%%%%%%%%%%%%%%%%%%%%%%
%%%%% figures
%%%%%%%%%%%%%%%%%%% %%%%%%%%%%%%%%%%%%%%%%%%%%%%%%%%%%%%%%%%%%%%%%
\newif\iffigureexists
\newif\ifepsfloaded
\def\epsfcheck{
\ifdraft% to speed up
\input epsf\epsfloadedtrue
\else
  \openin 1 epsf
  \ifeof 1 \epsfloadedfalse \else \epsfloadedtrue \fi
  \closein 1
  \ifepsfloaded
    \input epsf
  \else
\immediate\write20{NO EPSF FILE --- FIGURES WILL BE IGNORED}
  \fi
\fi
\def\epsfcheck{}}
\def\checkex#1{
\ifdraft
\figureexistsfalse\immediate%
\write20{Draftmode: figure #1 not included}
\figureexiststrue
\else\relax
    \ifepsfloaded \openin 1 #1
        \ifeof 1
           \figureexistsfalse
  \immediate\write20{FIGURE FILE #1 NOT FOUND}
        \else \figureexiststrue
        \fi \closein 1
    \else \figureexistsfalse
    \fi
\fi}
\def\missbox#1#2{$\vcenter{\hrule
\hbox{\vrule height#1\kern1.truein
\raise.5truein\hbox{#2} \kern1.truein \vrule} \hrule}$}
\def\lfig#1{%  this is to call the figure in the text
\let\labelflag=#1%
\def\numb@rone{#1}%
\ifx\labelflag\UnDeFiNeD%
{\xdef#1{\the\figno}%
\writedef{#1\leftbracket{\the\figno}}%
\global\advance\figno by1%
}\fi{\hyperref{}{figure}{{\numb@rone}}{Fig.{\numb@rone}}}}
\def\figinsert#1#2#3#4{%  this inserts the figure
\epsfcheck\checkex{#4}%
\def\figsize{#3}%
\let\flag=#1\ifx\flag\UnDeFiNeD
{\xdef#1{\the\figno}%
\writedef{#1\leftbracket{\the\figno}}%
\global\advance\figno by1%
}\fi
\goodbreak\midinsert%
\iffigureexists
\centerline{\epsfysize\figsize\epsfbox{#4}}%
\else%
\vskip.05truein
  \ifepsfloaded
  \ifdraft
  \centerline{\missbox\figsize{Draftmode: #4 not included}}%
  \else
  \centerline{\missbox\figsize{#4 not found}}
  \fi
  \else
  \centerline{\missbox\figsize{epsf.tex not found}}
  \fi
\vskip.05truein
\fi%
{\smallskip%
\leftskip 4pc \rightskip 4pc%
\noindent\ninepoint\sl \baselineskip=11pt%
{\bf{\hyperdef\hypernoname{figure}{{#1}}{Fig.{#1}}}:~}#2%
\smallskip}\bigskip\endinsert%
}

\def\boxit#1{\vbox{\hrule\hbox{\vrule\kern8pt
\vbox{\hbox{\kern8pt}\hbox{\vbox{#1}}\hbox{\kern8pt}}
\kern8pt\vrule}\hrule}}
\def\mathboxit#1{\vbox{\hrule\hbox{\vrule\kern8pt\vbox{\kern8pt
\hbox{$\displaystyle #1$}\kern8pt}\kern8pt\vrule}\hrule}}
%
%%%%%%%%%%%%%%%%%%% %%%%%%%%%%%%%%%%%%%%%%%%%%%%%%%%%%%%%%%%%%%%%%
%%%%%  fonts
%%%%%%%%%%%%%%%%%%% %%%%%%%%%%%%%%%%%%%%%%%%%%%%%%%%%%%%%%%%%%%%%%
\font\bigit=cmti10 scaled \magstep1

\font\titla=cmr10 scaled\magstep3
\font\tenmss=cmss10
\font\absmss=cmss10 scaled\magstep1

\newfam\mssfam
\font\footrm=cmr8  \font\footrms=cmr5
\font\footrmss=cmr5   \font\footi=cmmi8
\font\footis=cmmi5   \font\footiss=cmmi5
\font\footsy=cmsy8   \font\footsys=cmsy5
\font\footsyss=cmsy5   \font\footbf=cmbx8
\font\footmss=cmss8
\def\footfont{\def\rm{\fam0\footrm}
\textfont0=\footrm \scriptfont0=\footrms
\scriptscriptfont0=\footrmss
\textfont1=\footi \scriptfont1=\footis
\scriptscriptfont1=\footiss
\textfont2=\footsy \scriptfont2=\footsys
\scriptscriptfont2=\footsyss
\textfont\itfam=\footi \def\it{\fam\itfam\footi}
\textfont\mssfam=\footmss \def\mss{\fam\mssfam\footmss}
\textfont\bffam=\footbf \def\bf{\fam\bffam\footbf} \rm}
\def\tenpoint{\def\rm{\fam0\tenrm}
\textfont0=\tenrm \scriptfont0=\sevenrm
\scriptscriptfont0=\fiverm
\textfont1=\teni  \scriptfont1=\seveni
\scriptscriptfont1=\fivei
\textfont2=\tensy \scriptfont2=\sevensy
\scriptscriptfont2=\fivesy
\textfont\itfam=\tenit \def\it{\fam\itfam\tenit}
\textfont\mssfam=\tenmss \def\mss{\fam\mssfam\tenmss}
\textfont\bffam=\tenbf \def\bf{\fam\bffam\tenbf} \rm}
\ifx\answ\bigans\def\abstractfont{\tenpoint}\else
\def\abstractfont{\def\rm{\fam0\absrm}
\textfont0=\absrm \scriptfont0=\absrms
\scriptscriptfont0=\absrmss
\textfont1=\absi \scriptfont1=\absis
\scriptscriptfont1=\absiss
\textfont2=\abssy \scriptfont2=\abssys
\scriptscriptfont2=\abssyss
\textfont\itfam=\bigit \def\it{\fam\itfam\bigit}
\textfont\mssfam=\absmss \def\mss{\fam\mssfam\absmss}
\textfont\bffam=\absbf \def\bf{\fam\bffam\absbf}\rm}\fi
%
%%%%%%%%%%%%%%%%%%%%%%%%%%%%% %%%%%%%%%%%%%%%%%%%%%%%%%%%%%%%%%
%%%%% footnotes   (adapted from PHYZZX, no hypertext yet)
%%%%%%%%%%%%%%%%%%%%%%%%%%%%% %%%%%%%%%%%%%%%%%%%%%%%%%%%%%%%%%
\def\f@@t{\baselineskip10pt\lineskip0pt\lineskiplimit0pt
\bgroup\aftergroup\@foot\let\next}
\setbox\strutbox=\hbox{\vrule height 8.pt depth 3.5pt width\z@}
\def\vfootnote#1{\insert\footins\bgroup
\baselineskip10pt\footfont
\interlinepenalty=\interfootnotelinepenalty
\floatingpenalty=20000
\splittopskip=\ht\strutbox \boxmaxdepth=\dp\strutbox
\leftskip=24pt \rightskip=\z@skip
\parindent=12pt \parfillskip=0pt plus 1fil
\spaceskip=\z@skip \xspaceskip=\z@skip
\Textindent{$#1$}\footstrut\futurelet\next\fo@t}
\def\Textindent#1{\noindent\llap{#1\enspace}\ignorespaces}
\def\foot{\global\advance\ftno by1%
\attach{\hyperref{}{footnote}{\the\ftno}{\footsymbolgen}}%
\vfootnote{\hyperdef\hypernoname{footnote}{\the\ftno}{\footsymbol}}}%
%   this is for custom footnote marks:
\def\footnote#1{\global\advance\ftno by1%
\attach{\hyperref{}{footnote}{\the\ftno}{#1}}%
\vfootnote{\hyperdef\hypernoname{footnote}{\the\ftno}{#1}}}%
\newcount\lastf@@t           \lastf@@t=-1
\newcount\footsymbolcount    \footsymbolcount=0
\global\newcount\ftno \global\ftno=0
\def\footsymbolgen{\relax\footsym
\global\lastf@@t=\pageno\footsymbol}
\def\footsym{\ifnum\footsymbolcount<0
\global\footsymbolcount=0\fi
{\iffrontpage \else \advance\lastf@@t by 1 \fi
\ifnum\lastf@@t<\pageno \global\footsymbolcount=0
\else \global\advance\footsymbolcount by 1 \fi }
\ifcase\footsymbolcount
\fd@f\dagger\or \fd@f\diamond\or \fd@f\ddagger\or
\fd@f\natural\or \fd@f\ast\or \fd@f\bullet\or
\fd@f\star\or \fd@f\nabla\else \fd@f\dagger
\global\footsymbolcount=0 \fi }
\def\fd@f#1{\xdef\footsymbol{#1}}
\def\space@ver#1{\let\@sf=\empty \ifmmode #1\else \ifhmode
\edef\@sf{\spacefactor=\the\spacefactor}
\unskip${}#1$\relax\fi\fi}
\def\attach#1{\space@ver{\strut^{\mkern 2mu #1}}\@sf}
%
%%%%%%%%%%%%%%%%%%% %%%%%%%%%%%%%%%%%%%%%%%%%%%%%%%%%%%%%%%%%%%%%%
%%%%% References
%%%%%%%%%%%%%%%%%%% %%%%%%%%%%%%%%%%%%%%%%%%%%%%%%%%%%%%%%%%%%%%%%
\newif\ifnref
\def\rrr#1#2{\relax\ifnref\nref#1{#2}\else\ref#1{#2}\fi}
\def\ldf#1#2{\begingroup\obeylines
\gdef#1{\rrr{#1}{#2}}\endgroup\unskip}

\def\doubref#1#2{\refs{{#1},{#2}}}

\nreffalse
\def\refout{\listrefs}

\def\lref{\ldf}

%%%%%%%%%%%%%%%%%%% %%%%%%%%%%%%%%%%%%%%%%%%%%%%%%%%%%%%%%%%%%%%%%
%%%%%%% eq numbering
%%%%%%%%%%%%%%%%%%% %%%%%%%%%%%%%%%%%%%%%%%%%%%%%%%%%%%%%%%%%%%%%%
\def\eqn#1{\xdef #1{(\noexpand\hyperref{}%
{equation}{\secsym\the\meqno}%
{\secsym\the\meqno})}\eqno(\hyperdef\hypernoname{equation}%
{\secsym\the\meqno}{\secsym\the\meqno})\eqlabeL#1%
\writedef{#1\leftbracket#1}\global\advance\meqno by1}
\def\eqnalign#1{\xdef #1{\noexpand\hyperref{}{equation}%
{\secsym\the\meqno}{(\secsym\the\meqno)}}%
\writedef{#1\leftbracket#1}%
\hyperdef\hypernoname{equation}%
{\secsym\the\meqno}{\e@tf@ur#1}\eqlabeL{#1}%
\global\advance\meqno by1}
%old:
\def\eqnalign#1{\xdef #1{(\secsym\the\meqno)}
\writedef{#1\leftbracket#1}%
\global\advance\meqno by1 #1\eqlabeL{#1}}
%
%%%%%%%%%%%%%%%%%%% %%%%%%%%%%%%%%%%%%%%%%%%%%%%%%%%%%%%%%%%%%%%%%
%%%%%%  macros for titlepage, marginnotes, etc
%%%%%%%%%%%%%%%%%%% %%%%%%%%%%%%%%%%%%%%%%%%%%%%%%%%%%%%%%%%%%%%%%

%
\def\chap#1{\newsec{#1}}
\def\chapter#1{\chap{#1}}
\def\sect#1{\subsec{#1}}
\def\section#1{\sect{#1}}
\def\\{\ifnum\lastpenalty=-10000\relax
\else\hfil\penalty-10000\fi\ignorespaces}
\def\note#1{\leavevmode%
\edef\@@marginsf{\spacefactor=\the\spacefactor\relax}%
\ifdraft\strut\vadjust{%
\hbox to0pt{\hskip\hsize%
\ifx\answ\bigans\hskip.1in\else\hskip .1in\fi%
\vbox to0pt{\vskip-\dp
%\vskip4pt
\strutbox\sevenbf\baselineskip=8pt plus 1pt minus 1pt%
\ifx\answ\bigans\hsize=.7in\else\hsize=.35in\fi%
\tolerance=5000 \hbadness=5000%
\leftskip=0pt \rightskip=0pt \everypar={}%
\raggedright\parskip=0pt \parindent=0pt%
\vskip-\ht\strutbox\noindent\strut#1\par%
\vss}\hss}}\fi\@@marginsf\kern-.01cm}
\def\titlepage{%
\frontpagetrue\nopagenumbers\abstractfont%
\hsize=\hstitle\rightline{\vbox{\baselineskip=10pt%
{\abstractfont\pubnum}}}\pageno=0}
\frontpagefalse
\def\pubnum{}
\def\pdate{\number\month/\number\yearltd}
\def\makefootline{\iffrontpage\vskip .27truein
\line{\the\footline}
%\vskip -.1truein\line{\pdate\hfil}
\vskip -.1truein\leftline{\vbox{\baselineskip=10pt%
{\abstractfont\pdate}}}
\else\vskip.5cm\line{\hss \tenrm $-$ \folio\ $-$ \hss}\fi}
\def\title#1{\vskip .7truecm\titlestyle{\titleft #1}}
\def\titlestyle#1{\par\begingroup \interlinepenalty=9999
\leftskip=0.02\hsize plus 0.23\hsize minus 0.02\hsize
\rightskip=\leftskip \parfillskip=0pt
\hyphenpenalty=9000 \exhyphenpenalty=9000
\tolerance=9999 \pretolerance=9000
\spaceskip=0.333em \xspaceskip=0.5em
\noindent #1\par\endgroup }
\def\autskip{\ifx\answ\bigans\vskip.5truecm\else\vskip.1cm\fi}
\def\author#1{\vskip .7in \centerline{#1}}

\def\address#1{\ifx\answ\bigans\vskip.2truecm
\else\vskip.1cm\fi{\it \centerline{#1}}}
\def\abstract#1{
\vskip .5in\vfil\centerline
{\bf Abstract}\penalty1000
{{\smallskip\ifx\answ\bigans\leftskip 2pc \rightskip 2pc
\else\leftskip 5pc \rightskip 5pc\fi
\noindent\abstractfont \baselineskip=12pt
{#1} \smallskip}}
\penalty-1000}
\def\endpage{\tenpoint\supereject\global\hsize=\hsbody%
\frontpagefalse\footline={\hss\tenrm\folio\hss}}
\def\ack{\vskip2.cm\centerline{{\bf Acknowledgements}}}
%
%

%
%%%%%%%%%%%%%%%%%%%%%%%%%%%%% %%%%%%%%%%%%%%%%%%%%%%%%%%%%%%%%%
\def\bfone{\relax{\rm 1\kern-.35em 1}}
\def\inbar{\vrule height1.5ex width.4pt depth0pt}
\def\IC{\relax\,\hbox{$\inbar\kern-.3em{\mss C}$}}
\def\ID{\relax{\rm I\kern-.18em D}}
\def\IF{\relax{\rm I\kern-.18em F}}
\def\IH{\relax{\rm I\kern-.18em H}}
\def\II{\relax{\rm I\kern-.17em I}}
\def\IN{\relax{\rm I\kern-.18em N}}
\def\IP{\relax{\rm I\kern-.18em P}}
\def\IQ{\relax\,\hbox{$\inbar\kern-.3em{\rm Q}$}}
\def\IR{\relax{\rm I\kern-.18em R}}
\font\cmss=cmss10 \font\cmsss=cmss10 at 7pt
\def\ZZ{\relax\ifmmode\mathchoice
{\hbox{\cmss Z\kern-.4em Z}}{\hbox{\cmss Z\kern-.4em Z}}
{\lower.9pt\hbox{\cmsss Z\kern-.4em Z}}
{\lower1.2pt\hbox{\cmsss Z\kern-.4em Z}}\else{\cmss Z\kern-.4em
Z}\fi}
\def\a{\alpha}

\def\cC{{\cal C}}

 \def\cM{{\cal M}}
\def\cN{{\cal N}}

\def\nup#1({Nucl.\ Phys.\ $\us {B#1}$\ (}
\def\plt#1({Phys.\ Lett.\ $\us  {#1}$\ (}
\def\cmp#1({Comm.\ Math.\ Phys.\ $\us  {#1}$\ (}
\def\prp#1({Phys.\ Rep.\ $\us  {#1}$\ (}
\def\prl#1({Phys.\ Rev.\ Lett.\ $\us  {#1}$\ (}
\def\prv#1({Phys.\ Rev.\ $\us  {#1}$\ (}
\def\mpl#1({Mod.\ Phys.\ Let.\ $\us  {A#1}$\ (}
\def\ijmp#1({Int.\ J.\ Mod.\ Phys.\ $\us{A#1}$\ (}
\def\tit#1|{{\it #1},\ }
%
%%%%%%%%%%%%%%%%%%%%%%%%%%%%%%%% %%%%%%%%%%%%%%%%%%%%%%%%%%%%%%
%%%%% misc %%%%
%%%%%%%%%%%%%%%%%%%%%%%%%%%%%%%% %%%%%%%%%%%%%%%%%%%%%%%%%%%%%%

%

\def\ni{\noindent}

\def\us#1{\underline{#1}}

\def\hat{\widehat}

\def\Coe#1.#2.{{#1\over #2}}

\def\coe#1.#2.{\relax{\textstyle {#1 \over #2}}\displaystyle}
\def\half{{1 \over 2}}

\def\to{\rightarrow}
\def\notin{\hbox{{$\in$}\kern-.51em\hbox{/}}}

\def\del{\partial}

%%%%%%%%%%%%%%%%%%%%%%%%%%%

%%%%%%%%%%%%%%%%%%%%%%%%%%%
\catcode`\@=12
%%%%%%%%% end macros  %%%%%%% %%%%%%%%%%%%%%%%%%%%%%%%%%%%%%
%%%%%%%%%%%%%%%%%%%%%%%%%%%% %%%%%%%%%%%%%%%%%%%%%%%%%%%%%

%%%%%%%%%%%%%%%%%%%%%%%%%%%% %%%%%%%%%%%%%%%%%%%%%%%%%%%%%
% references: 
%%%%%%%%%%%%%%%%%%%%%%%%%%%% %%%%%%%%%%%%%%%%%%%%%%%%%%%%%

%\def\eprt#1{{\tt #1}}
\def\nihil#1{{\sl #1}}
\def\br{\hfill\break}

\def\ijmp {{Int. J. Mod. Phys.\ }{\bf A}}

\lref\arn{V.\ Arnold, S.\ Gusein-Zade and A.\ Varchenko,
{\it Singularities of Differentiable Maps},
Birkh\"auser, 1988.}

\lref\BDLR{I.\ Brunner, M.R.\ Douglas, A.\ Lawrence and 
C.\ R\"omelsberger, 
\nihil{D-branes on the quintic,}
\eprt{hep-th/9906200}. 
%%CITATION = HEP-TH 9906200;%%
}

\lref\OOY{H.\ Ooguri, Y.\ Oz and Z.\ Yin, 
\nihil{D-branes on Calabi-Yau spaces and their mirrors,}
 Nucl.\ Phys.\ {\bf B477} 407 (1996), 
\eprt{hep-th/9606112}. 
%%CITATION = HEP-TH 9606112;%%
}

\lref\gepnE{
D.\ Gepner, 
\nihil{Space-time supersymmetry in compactified string theory and 
superconformal models,}
 Nucl.\ Phys.\ {\bf B296} (1988) 757.
}
  
\lref\bound{
See e.g.,:
{A.\ Recknagel and V.\ Schomerus, 
\nihil{D-branes in Gepner models,}
 Nucl.\ Phys.\ {\bf B531} 185 (1998), 
\eprt{hep-th/9712186};\br 
%%CITATION = NUPHA,B531,185;%%
}
{J.\ Fuchs and C.\ Schweigert, 
\nihil{Branes: From free fields to general backgrounds,}
 Nucl.\ Phys.\ {\bf B530} 99 (1998), 
\eprt{hep-th/9712257};\br 
%%CITATION = HEP-TH 9712257;%%
}
{M.\ Gutperle and Y.\ Satoh, 
\nihil{D-branes in Gepner models and supersymmetry,}
 Nucl.\ Phys.\ {\bf B543} 73 (1999), 
\eprt{hep-th/9808080};\br
%%CITATION = NUPHA,B543,73;%%
}
{S.\ Govindarajan, T.\ Jayaraman and T.\ Sarkar, 
\nihil{World sheet approaches to D-branes on supersymmetric cycles,}
\eprt{hep-th/9907131};\br 
%%CITATION = HEP-TH 9907131;%%
}
{K.\ Hori and C.\ Vafa, 
\nihil{Mirror symmetry,}
\eprt{hep-th/0002222};\br 
%%CITATION = HEP-TH 0002222;%%
}
{K.\ Hori, A.\ Iqbal and C.\ Vafa, 
\nihil{D-branes and mirror symmetry,}
\eprt{hep-th/0005247}. 
%%CITATION = HEP-TH 0005247;%%
}
}

\lref\PGOH{P.\ A.\ Griffin and O.\ F.\ Hernandez, 
\nihil{Feigin-Fuchs derivation of SU(1,1) parafermion characters,}
 Nucl.\ Phys.\ {\bf B356} 287 (1991). 
%%CITATION = NUPHA,B356,287;%%
}

\lref\ASCV{A.\ D.\ Shapere and C.\ Vafa, 
\nihil{BPS structure of Argyres-Douglas superconformal theories,}
\eprt{hep-th/9910182}. 
%%CITATION = HEP-TH 9910182;%%
}

\lref\geomeng{
{S.\ Katz, A.\ Klemm and C.\ Vafa, 
\nihil{Geometric engineering of quantum field theories,}
 Nucl.\ Phys.\ {\bf B497} 173 (1997), 
\eprt{hep-th/9609239};\br 
%%CITATION = HEP-TH 9609239;%%
}
{S.\ Katz, P.\ Mayr and C.\ Vafa, 
\nihil{Mirror symmetry and exact solution of 4D N = 2 gauge theories.\ I,}
 Adv.\ Theor.\ Math.\ Phys.\ {\bf 1} 53 (1998), 
\eprt{hep-th/9706110}. 
%%CITATION = HEP-TH 9706110;%%
}
}

\lref\HIV{K.\ Hori, A.\ Iqbal and C.\ Vafa, as in \bound.}

\lref\GJ{S.\ Govindarajan and T.\ Jayaraman, 
\nihil{On the Landau-Ginzburg description of boundary CFTs and special 
Lagrangian submanifolds,}
\eprt{hep-th/0003242}. 
%%CITATION = HEP-TH 0003242;%%
}

\lref\SS{
{C.\ Vafa and N.\ Warner, 
\nihil{Catastrophes And The Classification Of Conformal Theories,}
 Phys.\ Lett.\ {\bf B218} 51 (1989);\br 
%%CITATION = PHLTA,B218,51;%%
}
{E.\ J.\ Martinec, 
\nihil{Criticality, Catastrophes And Compactifications,}
 Print-89-0373 (EFI,CHICAGO),{\it  In *Brink, L.\ (ed.) et al.: 
Physics and mathematics of strings* 389-433.}
}
}

\lref\otherCY{
{M.R.\ Douglas, 
\nihil{Topics in D-geometry,}
\eprt{hep-th/9910170};\br 
%%CITATION = HEP-TH 9910170;%%
}
{D.\ Diaconescu and C.\ R\"omelsberger, 
\nihil{D-branes and bundles on elliptic fibrations,}
\eprt{hep-th/9910172};\br 
%%CITATION = HEP-TH 9910172;%%
}
{P.\ Kaste, W.\ Lerche, C.\ A.\ L\"utken and J.\ Walcher, 
\nihil{D-branes on K3-fibrations,}
\eprt{hep-th/9912147};\br
%%CITATION = HEP-TH 9912147.%% 
%\href{\wwwspires?eprint=HEP-TH/9912147}{SPIRES}
}
{E.\ Scheidegger, 
\nihil{D-branes on some one- and two-parameter 
Calabi-Yau hypersurfaces,}
\eprt{hep-th/9912188};\br 
%%CITATION = HEP-TH 9912188;%% 
%\href{\wwwspires?eprint=HEP-TH/9912188}{SPIRES}
}
{I.\ Brunner and V.\ Schomerus, 
\nihil{D-branes at singular curves of Calabi-Yau compactifications,}
 JHEP{\bf 0004} 020 (2000), 
\eprt{hep-th/0001132};\br 
%%CITATION = HEP-TH 0001132;%%
}
{M.\ Naka and M.\ Nozaki, 
\nihil{Boundary states in Gepner models,}
 JHEP{\bf 0005} 027 (2000), 
\eprt{hep-th/0001037};\br 
%%CITATION = HEP-TH 0001037;%%
}
{K.\ Sugiyama, 
\nihil{Comments on central charge of topological 
sigma model with Calabi-Yau target space,}
\eprt{hep-th/0003166};\br 
%%CITATION = HEP-TH 0003166;%%
}
{M.\ R.\ Douglas, B.\ Fiol and C.\ R\"omelsberger, 
\nihil{The spectrum of BPS branes on a noncompact Calabi-Yau,}
\eprt{hep-th/0003263}. 
%%CITATION = HEP-TH 0003263;%%
}
}

\lref\GK{A.\ Giveon and D.\ Kutasov, 
\nihil{Little string theory in a double scaling limit,}
 JHEP{\bf 9910} 034 (1999), 
\eprt{hep-th/9909110}. 
%%CITATION = HEP-TH 9909110;%%
}

\lref\DFR{M.\ R.\ Douglas, B.\ Fiol and C.\ R\"omelsberger, 
\nihil{Stability and BPS branes,}
\eprt{hep-th/0002037}.
%%CITATION = HEP-TH 0002037;%%
}
\lref\DDCR{D.\ Diaconescu and C.\ R\"omelsberger, 
as in \otherCY.}        

\lref\OV{H.\ Ooguri and C.\ Vafa, 
\nihil{Two-Dimensional Black Hole and Singularities of CY Manifolds,}
 Nucl.\ Phys.\ {\bf B463} 55 (1996), 
\eprt{hep-th/9511164}. 
%%CITATION = HEP-TH 9511164;%%
}

\lref\DPL{L.\ J.\ Dixon, M.\ E.\ Peskin and J.\ Lykken, 
\nihil{N=2 Superconformal Symmetry And SO(2,1) Current Algebra,}
 Nucl.\ Phys.\ {\bf B325} 329 (1989). 
%%CITATION = NUPHA,B325,329;%%
}

\lref\OOY{H.\ Ooguri, Y.\ Oz and Z.\ Yin, 
\nihil{D-branes on Calabi-Yau spaces and their mirrors,}
 Nucl.\ Phys.\ {\bf B477} 407 (1996), 
\eprt{hep-th/9606112}. 
%%CITATION = HEP-TH 9606112;%%
}

\lref\KLM{A.\ Klemm, W.\ Lerche and P.\ Mayr, 
\nihil{K3 Fibrations and heterotic type II string duality,}
 Phys.\ Lett.\ {\bf B357} 313 (1995), 
\eprt{hep-th/9506112}. 
%%CITATION = HEP-TH 9506112;%%
}

\lref\KLMVW{A.\ Klemm, W.\ Lerche, P.\ Mayr, C.\ Vafa and N.\ Warner, 
\nihil{Self-Dual Strings and N=2 Supersymmetric Field Theory,}
 Nucl.\ Phys.\ {\bf B477} 746 (1996), 
\eprt{hep-th/9604034}. 
%%CITATION = HEP-TH 9604034;%%
}

\lref\curves{
{A.\ Klemm, W.\ Lerche, S.\ Yankielowicz and S.\ Theisen, 
\nihil{Simple singularities and N=2 supersymmetric Yang-Mills theory,}
 Phys.\ Lett.\ {\bf B344} 169 (1995), 
\eprt{hep-th/9411048};\br
%%CITATION = HEP-TH 9411048;%%
}
{P.\ C.\ Argyres and A.\ E.\ Faraggi, 
\nihil{The vacuum structure and spectrum of N=2 
supersymmetric SU(n) gauge theory,}
 Phys.\ Rev.\ Lett.\ {\bf 74} 3931 (1995), 
\eprt{hep-th/9411057};\br 
%%CITATION = HEP-TH 9411057;%%
}
{E.\ Martinec and N.\ Warner, 
\nihil{Integrable systems and supersymmetric gauge theory,}
 Nucl.\ Phys.\ {\bf B459} 97 (1996), 
\eprt{hep-th/9509161}. 
%%CITATION = HEP-TH 9509161;%%
}
}

\lref\KLTY{A.\ Klemm, W.\ Lerche, S.\ Yankielowicz and S.\ Theisen,
as in ref.\ \curves.}

\lref\AD{P.\ C.\ Argyres and M.\ R.\ Douglas, 
\nihil{New phenomena in SU(3) supersymmetric gauge theory,}
 Nucl.\ Phys.\ {\bf B448} 93 (1995), 
\eprt{hep-th/9505062}. 
%%CITATION = HEP-TH 9505062;%%
}

\lref\TH{T.\ J.\ Hollowood, 
\nihil{Strong coupling N = 2 gauge theory with arbitrary gauge group,}
 Adv.\ Theor.\ Math.\ Phys.\ {\bf 2} 335 (1998), 
\eprt{hep-th/9710073}. 
%%CITATION = HEP-TH 9710073;%%
}

\lref\KLT{A.\ Klemm, W.\ Lerche and S.\ Theisen, 
\nihil{Nonperturbative effective actions of 
N=2 supersymmetric gauge theories,}
 Int.\ J.\ Mod.\ Phys.\ {\bf A11} 1929 (1996), 
\eprt{hep-th/9505150}. 
%%CITATION = HEP-TH 9505150;%%
}

\lref\SWrev{W.\ Lerche, 
\nihil{Introduction to Seiberg-Witten theory and its stringy origin,}
 Nucl.\ Phys.\ Proc.\ Suppl.\ {\bf 55B} 83 (1997), 
\eprt{hep-th/9611190}. 
%%CITATION = HEP-TH 9611190;%%
}

\lref\spokes{
{P.\ Berglund, P.\ Candelas, X.\ De La Ossa, 
A.\ Font, T.\ Hubsch, D.\ Jancic and F.\ Quevedo, 
\nihil{Periods for Calabi-Yau and Landau-Ginzburg vacua,}
 Nucl.\ Phys.\ {\bf B419} 352 (1994), 
\eprt{hep-th/9308005}; \br 
%%CITATION = HEP-TH 9308005;%% 
%\href{\wwwspires?eprint=HEP-TH/9308005}
}
{P.\ Berglund, E.\ Derrick, T.\ Hubsch and D.\ Jancic, 
\nihil{On periods for string compactifications,}
 Nucl.\ Phys.\ {\bf B420} 268 (1994), 
\eprt{hep-th/9311143}. 
%%CITATION = HEP-TH 9311143;%% 
%\href{\wwwspires?eprint=HEP-TH/9311143}{SPIRES}
}
}

\lref\integ{
{P.\ Fendley, S.\ D.\ Mathur, C.\ Vafa and N.\ P.\ Warner, 
\nihil{Integrable Deformations And Scattering Matrices 
For The N=2 Supersymmetric Discrete Series,}
 Phys.\ Lett.\ {\bf B243} 257 (1990);\br
%%CITATION = PHLTA,B243,257;%%
}
{P.\ Fendley, W.\ Lerche, S.\ D.\ Mathur and N.\ P.\ Warner, 
\nihil{N=2 supersymmetric integrable models from affine toda theories,}
 Nucl.\ Phys.\ {\bf B348} 66 (1991). 
%%CITATION = NUPHA,B348,66;%%
}
}

\lref\SW{N.\ Seiberg and E.\ Witten, 
\nihil{Electric - magnetic duality, 
monopole condensation, and confinement in N=2 
supersymmetric Yang-Mills theory,}
 Nucl.\ Phys.\ {\bf B426} 19 (1994), 
\eprt{hep-th/9407087}. 
%%CITATION = HEP-TH 9407087;%%
}

\lref\BF{
{F.\ Ferrari and A.\ Bilal, 
\nihil{The Strong-Coupling Spectrum of the Seiberg-Witten Theory,}
 Nucl.\ Phys.\ {\bf B469} 387 (1996), 
\eprt{hep-th/9602082};\br 
%%CITATION = HEP-TH 9602082;%%
}
{A.\ Bilal and F.\ Ferrari, 
\nihil{Curves of Marginal Stability and 
Weak and Strong-Coupling BPS Spectra in{$N=2$} Supersymmetric{QCD}}
 Nucl.\ Phys.\ {\bf B480} 589 (1996), 
\eprt{hep-th/9605101}. 
%%CITATION = HEP-TH 9605101;%%
}
}

\lref\DHT{
{N.\ Dorey, 
\nihil{The BPS spectra of two-dimensional 
supersymmetric gauge theories with twisted mass terms,}
 JHEP{\bf 9811} 005 (1998), 
\eprt{hep-th/9806056};\br 
%%CITATION = HEP-TH 9806056;%%
}
{N.\ Dorey, T.\ J.\ Hollowood and D.\ Tong, 
\nihil{The BPS spectra of gauge theories in two and four dimensions,}
 JHEP{\bf 9905} 006 (1999), 
\eprt{hep-th/9902134}. 
%%CITATION = HEP-TH 9902134;%%
}
}

\lref\CV{S.\ Cecotti and C.\ Vafa, 
\nihil{On classification of N=2 supersymmetric theories,}
 Commun.\ Math.\ Phys.\ {\bf 158} 569 (1993), 
\eprt{hep-th/9211097}. 
%%CITATION = HEP-TH 9211097;%%
}

%%%%%%% paper  specific macros

\def\N{{\cN}}
\def\IG{\relax\,\hbox{$\inbar\kern-.3em{\mss G}$}}

%%%%%%%%%%%%%%%%%%%%%%%%%%%% %%%%%%%%%%%%%%%%%%%%%%%%%%%%%

%\draft

\def\pubnum{
\hbox{CERN-TH/2000-163}
\hbox{hep-th/0006100}
\hbox{}}
\def\pdate{}
\titlepage
\vskip2.cm
\title{{\titlefont On a Boundary CFT Description of
 Nonperturbative $N=2$ Yang-Mills Theory}}
%\vskip -.7cm
\autskip
\author{\wl} \vskip.2truecm
\CERN
\vskip.2truecm

\abstract{
We describe a simple method for determining the strong-coupling BPS
spectrum of four dimensional $\N=2$ supersymmetric Yang-Mills theory.
The idea is to represent the magnetic monopoles and dyons in terms of
$D$-brane boundary states of a non-compact $d=2$ $\N=2$ Landau-Ginzburg
model. In this way the quantum truncated BPS spectrum at the origin  of
the moduli space can be directly mapped to the finite number of primary
fields of the superconformal minimal models.
}

\vfil
%\vskip 1.cm
\ni {CERN-TH/2000-163}\hfill\break
\ni June 2000
\endpage
\baselineskip=14pt plus 2pt minus 1pt

\sequentialequations

%%%%%%%%%%%%%%%%%%%%%%%%%%%%%%%%%%%%%%%%%%%%%
\chapter{Introduction}
%%%%%%%%%%%%%%%%%%%%%%%%%%%%%%%%%%%%%%%%%%%%%

Boundary conformal field theory (BCFT)  has turned out to be a very
useful tool for investigating properties of $D$-branes, especially in
the domain of strong quantum corrections. Most notably boundary $\N=2$
minimal models \bound\ have provided important results on the spectrum of
$D$-branes at the Fermat (``Gepner'') point of Calabi-Yau threefold
compactifications \doubref\BDLR\otherCY.

A particularly interesting feature of four dimensional theories with
$\N=2$ supersymmetry is that the $D$-brane BPS spectrum at the Gepner
point can be substantially different as compared to the large radius
limit, where semi-classical geometry applies. Indeed, when
interpolating between these regimes in the moduli space, one may cross
certain lines of marginal stability. On these certain central charges
become collinear and thus BPS states become unstable against decay
into constituents with smaller charges.

This phenomenon is well-known from $\N=2$ $d=2$ LG theory \CV, and in
the context of $d=4$ gauge theory it  had been first  discussed in
\refs{\SW,\BF}. While most investigations have been centered at gauge
group $G=SU(2)$,  there is only limited knowledge for general gauge
groups and matter content, 
because the situation becomes rapidly very complex. 
The picture that seemed
to emerge for pure gauge theories is that in a sufficiently small
neighborhood of the origin of the moduli space, $\cM_G$, the spectrum
of stable BPS states is finite and consists only of those monopoles and
dyons which become massless at the various singularities in $\cM_G$. On
the other hand, in the semi-classical regime at ``infinity'' in
$\cM_G$,  the BPS spectrum is infinite, its most prominent members
being the massive gauge fields.

The purpose of the present note is to determine the
strong coupling spectrum of $\N=2$ $SU(N)$ gauge theories, 
by relating them to
boundary Landau-Ginzburg theories \doubref\GJ\HIV\ and so effectively
mapping them to $\N=2$ $d=2$ minimal models at levels $k=N-2$. In this
way, the BPS states at the origin of $\cM_G$ can be mapped via boundary
states to the $N(N-1)$ primary fields $\phi^\ell_{m}$,  so that the
quantum truncation of the spectrum may be interpreted in terms of the
finite number of the primary fields in the 2d CFT.  It will turn out
that the stable BPS states have indeed precisely the electric and
magnetic RR charges of the potentially massless monopoles and dyons.

Our setup is very simple. As is well-known, $\N=2$ gauge theories can
be systematically embedded into type II string compactifications on
Calabi-Yau threefolds \doubref\KLMVW\geomeng. Specializing to the gauge
sub-sector in question amounts to focusing on a (neighborhood of an)
appropriate isolated singularity on the threefold. This can be modeled
in terms of a non-compact CY threefold, whose compact piece supports
the geometry of the relevant Seiberg-Witten curve. The monopoles and
dyons then correspond to $D$-branes wrapped around the compact cycles,
while the non-compact directions subsume the non-universal degrees of
freedom which decouple in the rigid theory we are interested in.

Concretely, for $G=SU(N)$ the non-compact threefold can be 
written as~\KLMVW:
$$
z+{1\over z}+P_{A_{N-1}}(x_1,u_k)+{x_2}^2+{x_3}^2\ =\ 0\ ,
\eqn\nonCY
$$ 
where $P_{A_{N-1}}(x,u_k)=x^N-\sum_{k=2}^N u_k x^{N-k}$ is the
normal form of the simple singularity \arn\ of type $A_{N-1}$;
the other simply laced gauge groups of type $D,E$ 
can be treated analogously. 
Dropping the quadratic pieces, \nonCY\ turns precisely into
the Riemann surface that underlies the BPS dynamics of
$\N=2$ Yang-Mills theory \doubref\SW\curves.
We will be interested in the ``Gepner point'' $u_k\equiv0$ of this 
geometry and study the spectrum of 
wrapped $D$-branes in terms of boundary CFT.

%%%%%%%%%%%%%%%%%%%%%%%%%%%%%%%%%%%%%%%%%%%%%%%%%%%%%%%
\chapter{Non-compact Landau-Ginzburg description}
%%%%%%%%%%%%%%%%%%%%%%%%%%%%%%%%%%%%%%%%%%%%%%%%%%%%%%%

While the form \nonCY\ of the non-compact threefold
is convenient for studying the
geometry that underlies the Yang-Mills theory, it is not
very useful for a CFT formulation, because for this we need an
$\N=2$ superconformal Landau-Ginzburg theory with $\hat c=3$ to start
with. In order to find a suitable form, recall that the geometry 
described by \nonCY\ is given by the fibration of an ALE space
(described here by $P_{A_{N-1}}$ besides the un-important quadratic
pieces) over a $\IP^1$ base (described by the $z$-dependent part) \KLMVW.
A good starting point is, therefore, to first focus on the ALE space.

The ALE space corresponds to a non-compact model of a type $A_{N-1}$
singularity on a compact $K3$ surface, and type II string
compactification on it can be described in terms of a CFT based on the
following Landau-Ginzburg superpotential \OV:\foot{We drop quadratic
terms because they are irrelevant for the LG theory.}
$$
W^{ALE}_{A_{N-1}}(x,z,u_k)\ =\ 
x^N-\sum_{k=2}^N u_k\, x^{N-k}z^{-k}\ ,
\eqn\nonALE
$$
and analogously for the other simply laced groups.
The theory is singular at $u_k=0$, and this reflects the 
appearance of massless gauge fields (or non-critical strings, 
if we start from the IIB theory instead of type IIA).
The Gepner point of this theory is described by
$$
W^{ALE}_{A_{N-1}}(x,z,u_N=-1)=x^N+{1\over z^N}\ ,
\eqn\ALEGep
$$
which indeed describes a smooth CFT with $\hat c=2$ (remember that each
term $x^N$ contributes $\hat c(N)=(N\!-\!2)/N$). It can also be
described in terms of a coset CFT based on  $\left({SU(2)_{N-2}\over
U(1)}\times {SL(2)_{N+2}\over U(1)}\right)/\ZZ_N$ \doubref\OV\GK.  The
idea is that the non-compact part of the theory is a placeholder that
encodes the non-universal, but decoupled dynamics that is irrelevant to
our problem. The quantities we are interested in reside  in the compact
sub-sector, and do not depend on the details of the non-compact CFT.

We now return to describing the geometry we are really after, namely a
fibration of the ALE space over $\IP^1$. As is familiar from compact
$K3$ fibrations \KLM, this ultimately amounts to splitting $z$ into two
coordinates with half the degree each, and so we arrive, tentatively at
first, at the following LG representation of the  Seiberg-Witten
theory:
$$
W^{SW}_{A_{N-1}}(x,z_1,z_2,u_k)\ =\ 
x^N+{1\over {z_1}^{2N}}+ {1\over {z_2}^{2N}}-\sum_{k=2}^N u_k\, x^{N-k}(z_1z_2)^{-k}.
\eqn\nonSW
$$
This describes a CFT with $\hat c=3$ which is smooth at the origin of
its moduli space, $u_k=0$; on the other hand, $u_N\to\infty$
corresponds to the large base $\IP^1$ limit where we recover the ALE
space. Moreover the singularities at $u_\ell=0,u_N=\pm2$, where the
purely $z$-dependent piece forms a complete square, correspond to
the  Argyres-Douglas points \AD.\foot{
In \ASCV\ a non-compact LG description near $SU(N)$ Argyres-Douglas
points has been used to analyze the BPS spectrum
near these points.} More generally one can check that the
discriminant locus in the moduli space is the same as for \nonCY. We
will thus take \nonSW\ as the defining superpotential for our boundary
Landau-Ginzburg theory.\foot{Indeed, changing variables like in \KLMVW\
and fixing a patch reproduces \nonCY.}

Our purpose is now to determine the spectrum of $A$-type \OOY\
boundary states at the Gepner point of \nonSW,
by focusing on the compact piece of
$$
W^{SW}_{A_{N-1}}(x,z_1,z_2)\ =\ 
x^N+{1\over {z_1}^{2N}}+ {1\over {z_2}^{2N}}\ .
\eqn\SWGep
$$
This may also be viewed as a coset model based on
$\left(\!{SU(2)_{N-2}\over
U(1)}\!\times\! \big({SL(2)_{2N+2}\over 
U(1)}\big)^2\!\right)\!/\ZZ_{2N}$.

%%%%%%%%%%%%%%%%%%%%%%%%%%%%%%%%%%%%%%%%%%%%%%%%%%%%%%%
\chapter{Boundary CFT and Intersection Index}
%%%%%%%%%%%%%%%%%%%%%%%%%%%%%%%%%%%%%%%%%%%%%%%%%%%%%%%

An important quantity to compute in order to verify that \SWGep\ is the
correct form of the Landau-Ginzburg potential is the topological
``intersection index'' $I_{a,b}\equiv {\rm Tr}_{a,b}[(-1)^F]$ of boundary
states $a$,$b$.  Using standard BCFT technology, the index will indeed
turn out to coincide with the intersection matrix of the {\it vanishing
cycles} (and not just of some arbitrary homology cycles) of the
Seiberg-Witten curve. Subsequently, we will determine the spectrum of
the stable wrapped $D$-branes.

Since the LG potential \SWGep\ represents a tensor product of $\N=2$
superconformal models, we can most conveniently focus on its
components. First of all, as is well known, $x^N$ represents an $\N=2$
minimal model $(SU(2)_{k}\times U(1))/ U(1)$ at level $k=N-2$. The primary
fields are labelled by $(\ell,m,s)$, with $\ell=0,...,N-2$,
$m=-N+1,\dots,N$ (mod $2N$), and in addition $s=-1,0,1,2$ (mod $4$)
determines the R- or NS-sectors $(\ell+m+s=0$ mod $2$). 
We will be interested in A-type 
boundary states $|\ell_i,m_i,s_i\rangle$, which are labelled
by the same letters as  the primary fields. 
As has been shown in recent papers \doubref\BDLR\HIV, 
the intersection index can be written as the following
overlap amplitude:
$$
I_{{\ell_1},{\ell_2}}(m_1,m_2,s_1,s_2)\ \equiv\
{\phantom{\big|}}_{{\rm RR}}\big\langle \ell_1,m_1,s_1\big|
\ell_2,m_2,s_2\big\rangle_{{\rm RR}}
%\ =\ 
%\Trbel{{(\ell_1,m_1,s_1),\atop (\ell_2,m_2,s_2)}}\big[(-1)^F\big]
\ .\eqn\defI
$$
Using the expansion of the boundary states into Ishibashi states
$|\ \rangle\rangle$,
$$
\big|\ell,m,s\big\rangle\ =\ \sum_{(\ell',m',s')}
{
S_{(\ell,m,s)}^{\ \ (\ell',m',s')}
\over
\sqrt{{S_{(0,0,0)}^{\ \ (\ell',m',s')}}}
}\,
\big|\ell',m',s' \big\rangle\big\rangle\ ,
\eqn\Ishi
$$
where
$
S_{(\ell,m,s)}^{\ \ (\ell',m',s')}={1\over\sqrt2N}
\sin\big[{\pi\over N}(\ell+1)(\ell'+1)\big]
\exp\big[i{\pi\over N}(mm'-{N\over2}ss')\big]
$
is the modular transformation matrix associated with
the $\N=2$ minimal model characters, the result is \doubref\BDLR\HIV:
$$
\big(I_{{\ell_1},{\ell_2}}\big)_{m_1}^{\ \ m_2}(s_1,s_2)\ =\
(-1)^{{s_2-s_1\over2}}\,N_{{\ell_1},{\ell_2}}^{m_2-m_1}\ .
\eqn\IeqN
$$
This can be considered as $2N\times 2N$ matrix for fixed $\ell_i$, $s_i$
(in the following, we will keep $s_i$ fixed). Above,
$$
N_{{\ell_1},{\ell_2}}^{\ell_3} =
{2\over N}\sum_{\ell=0}^k
{
\sin\big[{\pi\over N}(\ell_1+1)(\ell+1)\big]
\sin\big[{\pi\over N}(\ell_2+1)(\ell+1)\big]
\sin\big[{\pi\over N}(\ell_3+1)(\ell+1)\big]
\over
\sin\big[{\pi\over N}(\ell+1)\big]
}\eqn\FR
$$
are the Verlinde fusion coefficients associated with $SU(2)_k$.
Moreover one can extend the standard range of the upper index, 
by defining  $N_{{\ell_1},{\ell_2}}^{-\ell_3-2}\equiv -N_{{\ell_1},{\ell_2}}^{\ell_3}$
and $N_{{\ell_1},{\ell_2}}^{-1}=N_{{\ell_1},{\ell_2}}^{N-1}\equiv0$.
The extended fusion coefficients are then periodic
and so can be compactly written in terms of the
$\ZZ_N$ step generator $g(2N)\equiv \gamma^2(2N)$, where
$$
\gamma(K)  =  \left( 
\matrix{ 0 & 1 & 0 & \cdots & 0 & 0 \cr
         0 & 0 & 1 & \cdots & 0 & 0 \cr
         \vdots & \vdots & \vdots & \ddots & \vdots & \vdots \cr
         0 & 0 & 0 & \cdots & 0 & 1  \cr
     1 & 0 & 0 & \cdots & 0 & 0} \right)_{K\times K}\ . 
\eqn\shift
$$
One has then explicitly for the $\ell=0$ states \BDLR:
$$
I_{0,0}^{A_{N-1}}\ =\ {\bf 1}-g(2N)\ ,
\eqn\IAnn
$$
which is a matrix without definite symmetry properties. 
Due to the selection rule $l+m+s=0$ (mod $2$), only every other
entry needs to be considered in a given R- or NS-sector,
and this is what we will do when we write down
explicit matrices further below.

In order to obtain geometrically meaningful intersection matrices
associated with non-compact Calabi-Yau $\hat c-$folds, one needs to
augment \IAnn\ by a contribution of the non-compact sector which pushes
the central charge up to $\hat c=2$ or $\hat c=3$. This will ensure a
completely symmetric or anti-symmetric intersection matrix. Note
however that  the structure of the non-compact models $SL(2)_{N+2}\over
U(1)$ is much more complicated than the one of the minimal models. In
particular the $SL(2)$ fusion rules,  if they are well-defined at
all, are not known and do not even truncate, and it would be pretty
pointless to try to solve the associated boundary CFT.

Fortunately, this is not necessary and all we will need from the
non-compact sector is the intersection matrix for the trivial
representation, $\ell=0$. By choosing a parafermionic representation of
the $SU(2)_k$ Kac-Moody algebra ($k\!=\!N\!-\!2$), 
it is easy to see that when continuing
to negative $k$, which corresponds 
to going to $SL(2)_k$, the $U(1)$ current
$J^3=i\sqrt{k}\del \phi$ switches sign and the r\^ oles of $J^+$ and
$J^-$ are exchanged \doubref\DPL\PGOH. As a consequence primary fields
associated with parafermions $\psi^{\ \ell}_m$ will now be associated with
negative charges, $q=-m/N$.  Thus, while we certainly do not know the
exact $SL(2)_{N+2}\over U(1)$ $S$-matrices in the expansion of the
boundary states into Ishibashi states \Ishi,  the structure should
remain simple for the $\ell=0$ states as far as the labels $m$ are
concerned, and be related to the minimal model matrices by a sign flip
of the $m$ labels.

In other words,  $m_1$ and $m_2$ are exchanged in \IeqN\ and this just
amounts to the transposition of $I_{0,0}$. Imposing periodicity as
before, we then conclude that for a LG theory with $W=1/x^N$ the
intersection index should be:
$$
I_{0,0}\ =\ {\bf 1}-g^{-1}(2N)\ .
\eqn\Inn
$$
This is the generalization to negative $N$ of the rule
that in a tensor product LG model with $\ZZ_K$ scaling symmetry, 
each term $x^N$ contributes a factor $({\bf 1}-g^{K/N}(2K))$
to the intersection index \doubref\BDLR\DDCR. 
The rule can be understood also from geometry: $({\bf 1}-g^{K/N})$ 
is nothing but the ``variation operator'' that maps relative
to absolute homology \arn . It in particular appears 
when evaluating period integrals in terms of non-compact, V-shaped
integration contours \spokes.

In forming the tensor product, we still need for the GSO
projetion to mod out the overall $R$ symmetry, 
and this identifies the charge labels
of the individual factors. Thus all-in-all
we obtain from \ALEGep\ the following intersection
index for $D$-brane boundary states:
$$
(I_{0,0})^{ALE}_{A_{N-1}}\ =\ \Big({\bf 1}-g(2N)\Big)\cdot\Big({\bf 1}-g^{-1}(2N)\Big)\ .
\eqn\IALE
$$
Similarly we get from the Landau-Ginzburg potential \SWGep\
for the non-compact threefold:
$$
I_{0,0}^{SW}\ =\ 
\Big({\bf 1}-g^2(4N)\Big)
\cdot
\Big({\bf 1}-g^{-1}(4N)\Big)^2\ .
\eqn\ISW
$$
We will show in the next section that this is indeed the correct, fully
anti-symmetric intersection matrix of vanishing cycles of the
Seiberg-Witten curve. 

Before doing so we like to recall
that the intersection index for boundary states\foot{
Of course we mean here the higher $\ell$ states of the
minimal model only, and restrict ourselves to the 
$\ell=0$ sector of the non-compact sector.} with $\ell\not=0$
can be obtained from $I_{0,0}$ by simple matrix multiplication.
More specifically, all we need is to consider $\ell=0,...,[k/2]$,
because of the field identification $\phi^{\ \ell}_{m,s}=\phi^{\ k-\ell}_{m+k+2,s+2}$ we can map $\ell=[k/2]+1,...,k,$ back into 
the smaller range (the shifted $s$-label implies that these 
states correspond to the anti-branes; the exception is at the fixed point
$\ell=k/2$ where branes and anto-branes sit in the same $m$-orbit.).
From the fusion coefficients \FR\ one can then deduce \DDCR:
$$
\eqalign{
I_{\ell_1,\ell_2}\ &=\ t_{\ell_1}\cdot I_{0,0}\cdot t^t_{\ell_2}\ ,\cr
t_{\ell}\ &\equiv\ \sum_{k=-\ell/2}^{\ell/2}\gamma^{2k}\ ,
}\eqn\Itt
$$
where $\gamma$ is the square--root \shift\ of $g$.
It follows that if ${q_{(0)}}^i=\{0,0,..,1_i,0,..0\}$
are the charges of the ``basic'' boundary states with $\ell=0$,
the charges of the other states are 
$$
\vec q_{(\ell)}=\vec q_{(0)}\cdot t_{\ell}\ .
\eqn\charges
$$

%%%%%%%%%%%%%%%%%%%%%%%%%%%%%%%%%%%%%%%%%%%%%%%%%%%%%%%
\chapter{$D$-brane spectrum}
%%%%%%%%%%%%%%%%%%%%%%%%%%%%%%%%%%%%%%%%%%%%%%%%%%%%%%%

We now like to verify that the boundary states, whose
intersection indices are given by \IALE\ and \ISW,
indeed correspond to wrapped $D$-branes that represent gauge fields
and $\N=2$ SYM dyons, respectively. 

We start with the simpler case \IALE, where is easy to see  that the
submatrix obtained by extracting every other entry coincides with the
(cyclically extended) symmetric rank $(N-1)$ Cartan matrix
$\cC_{A_{N-1}}$ of $SU(N)$.\foot{The deeper reason is the fact
that the modular $S$ matrices which enter in the fusion rule
coefficients $N_{\ell_1,\ell_2}^{\ell_3}$ happen to be the eigenvectors
of the  $A_{N-1}$ Cartan matrix.}  Therefore, the $\ell=0$ boundary
states correspond to the simple roots $\a_i$, which means that one can
choose a basis  in which the boundary states have the following
charges:
$$
\vec q_{(0)}\ =\ \pmatrix{\a_1\cr\vdots\cr\a_{N-1}\cr\a_N}\ ,
\eqn\rootcharges
$$
where the extending root is defined by $\a_N\equiv-\sum^{N-1}_i\a_i$.
This is precisely as required if we want
to interpret the boundary states in terms of $D$-branes wrapped around
the compact $2$-cycles of the ALE space. It is well-known that these
cycles correspond to root vectors and intersect in a Dynkin
diagram-like pattern. 

However, in order that the $D$-branes describe gauge fields in the
adjoint representation, we will need not just the simple roots but all
the roots. One can easily see that the remaining roots are obtained
from the boundary states with higher spins, $\ell=1,...,[k/2]$. A
simple way to deduce this is to visualize a circle of $N$ points in the
$x$-plane, which can be thought of as the projection of a weight
diagram. The boundary states with $\ell=0$ then correspond to the
simple roots that connect subsequent points around the circle. Moreover
the $\ell=1$ states connect every other point, and so on.  Group
theoretically this corresponds to decomposing the adjoint
representation into orbits of the Coxeter element of the Weyl group. 
In total we obtain
all the $\half N(N-1)$ positively charged gauge fields in this way,
whose charges are given by linear combinations of the simple roots
precisely as given in \charges.

We now turn to the more interesting BCFT based on  the non-compact
threefold \SWGep. Geometrically, what happens when we fiber an ALE
space over a $\IP^1$ base is that each point in the $x$-plane splits
into a pair of branch points, and these are precisely the branch points
of the Seiberg-Witten curve \KLMVW\ (this has been discussed at length
in the literature, see e.g., \SWrev). Correspondingly each 2-cycle on
the ALE space splits into two 3-cycles on the 3-fold, which correspond
to 1-cycles on the SW curve; examples for such cycles are shown in Fig.1.

\figinsert\graphs{
On the left we see a representation of the  Seiberg-Witten curve for
$G=SU(6)$ at the origin of its moduli space. The fat lines
denote the branch cuts of a double cover of the $x$-plane. The dashed
lines denote the vanishing cycles that correspond to the
stable dyons (we show only the upper sheet). 
On the right we display the
intersection matrices of these cycles, which coincide with the
intersection indices  $I^{SW}_{\ell,\ell}$ \Itt\
of the boundary states with spin~$\ell$.
 }{3.5in}{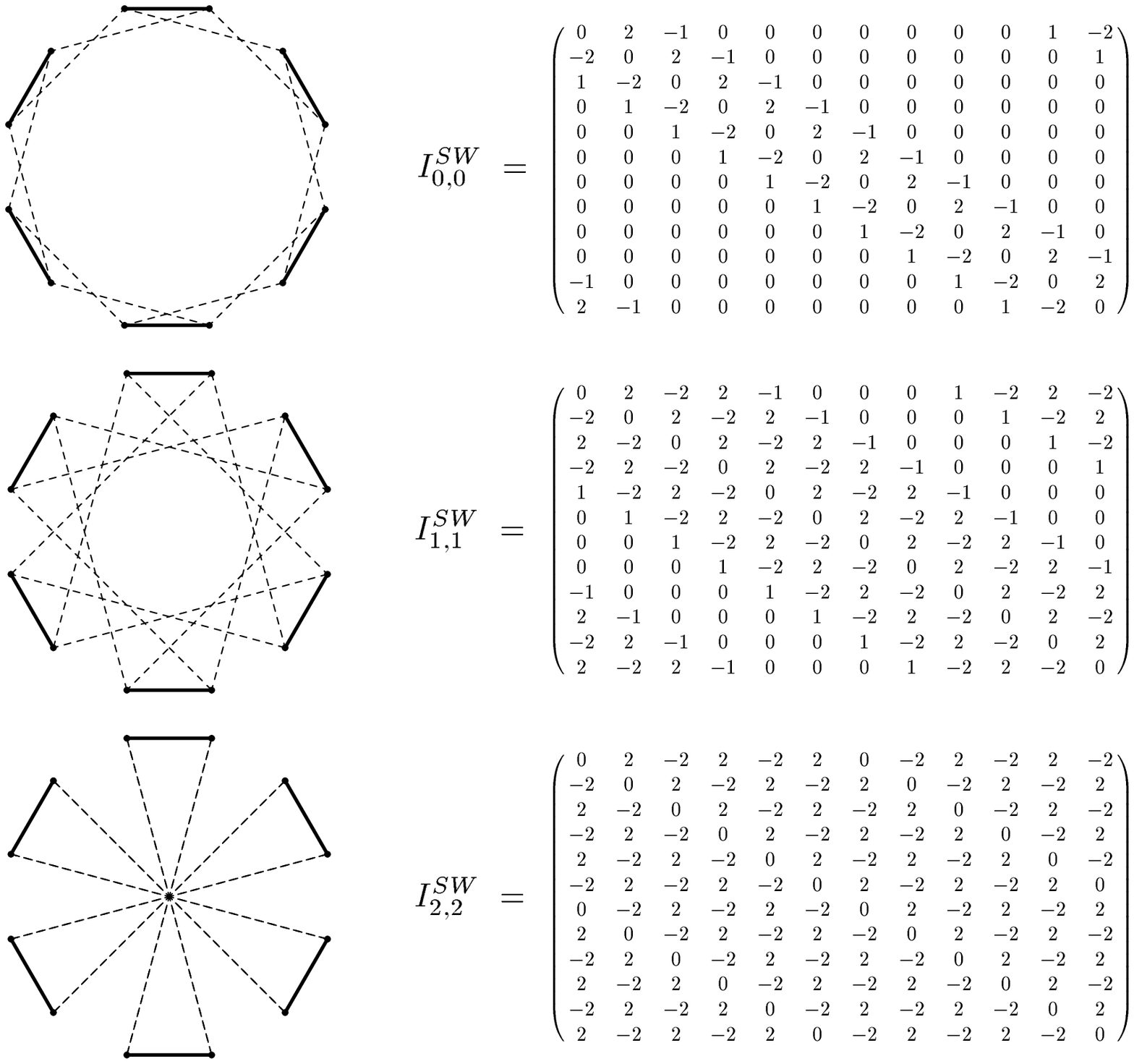}

A suitable symplectic basis of these cycles, 
which exhibits the magnetic and electric
charges in the form $q_i=[\, g_i\,;\,e_i\,]$,
can be chosen as follows:
$$
\vec q_{(0)}\ =\ 
\pmatrix{
[\ \ \ \ \a_1\ \ ;\ \ 0\ \ \ \ ] \cr
[\, \ \ -\a_1\ ;\ \a_1 \ \ \ ] \cr
\ \ \vdots\ \ \ \ \ \ \vdots \cr
[\ \ \ \a_i\ ;(i-1)\a_i ] \cr
[\, \ -\a_i\ ;(2-i)\a_i \ ] \cr
\ \ \vdots\ \ \ \ \ \ \vdots \cr
[\ \ \a_N\ ;\sum(1-k)\a_k] \cr
[\, -\a_N\ ;\sum(k-2)\a_k] \cr
}\ .
\eqn\SWchargevector
$$
These are precisely
the RR charge vectors of some the strong coupling dyons (namely
those which are associated with the simple roots)
\refs{\KLT,\SWrev,\TH}. The inner product metric of this basis 
takes a symplectic DZW form:
$$
(I_{geom}^{SW})_i^{\ j}\ =\ q_i\circ q_j \equiv\langle g_i,e_j\rangle
-\langle g_j,e_i\rangle\ ,
\eqn\Igeo
$$
where $\langle...,...\rangle$ is the inner product in weight space
which given by the Cartan matrix of $SU(N)$. 
$I_{geom}^{SW}$ is known to be the
geometric intersection matrix  of the vanishing cycles that correspond
to the potentially massless dyons \KLT. As one can easily verify, it
indeed coincides with the intersection index \ISW\ of the $\ell=0$
boundary states:
$$
I_{geom}^{SW}\ =\ I_{0,0}^{SW}\ .
$$
Moreover, there are boundary states with spins
$\ell=1,...,[k/2]$ (remember that $\ell=[k/2]+1,...,k$, corresponds
to the anti-branes for which the orientation of the cycles is
reversed). Similarly to what we have discussed for the ALE space, these
correspond in the Yang-Mills theory to the dyons that originate from
the non-simple roots. In total we have $2\cdot\half N(N-1)$ states
whose charges are determined by \charges, and one can check that these
charges match the corresponding cycles on the SW curve --  see again
Fig.1.

Finally some remarks on the BPS nature of the boundary  states,
following \HIV. Note that maximally rank$G=N-1$ of the dyons can be
mutually local with respect to each other, which corresponds to a
choice of $N-1$ non-intersecting cycles \AD\ (parallel dashed lines in
Fig.1). Any given choice of such boundary states can be mapped to the
set of primary chiral fields $\phi_{\ell}^{\ \ell}$ of the superconformal
mininal model at level $k=N-2$. They are BPS with respect to the same
linear combination of left- and right-moving supercharges. The other
possible choices of mutually local states are generated by the
$\ZZ_{2N}$ symmetry, which describe states that preserve different
linear combinations of the supercharges \HIV. Altogether the $N(N-1)$
states (not counting the anti-branes) exhaust the set of primary (not
necessarily chiral) fields $\phi_{m}^{\ \ell}$ in the minimal model.

Summarizing, we find that applying the rules of boundary conformal
field theory to the non-compact LG potential \SWGep,
we find a complete match between the charges of the boundary states
and the charges of those Yang-Mills dyons that are supposedly
stable at the origin of the moduli space.

%%%%%%%%%%%%%%%%%%%%%%%%%%%%%%%%%%%%%%%%%%%%%%%%%%%%%%%
\chapter{Discussion}
%%%%%%%%%%%%%%%%%%%%%%%%%%%%%%%%%%%%%%%%%%%%%%%%%%%%%%%

The important point of the present paper is not so much that the
intersection indices $I_{\ell,\ell}={\rm Tr}_{\ell,\ell}[(-1)^F]$, as
computed from CFT fusion rules, coincide with the geometric
intersection matrices of the vanishing cycles on the ALE space or SW
curve. This is just a reflection of the universality of the simple
singularities, which happen to underlie both the Yang-Mills theory
\KLTY\ and the superconformal minimal models \SS.  At least we may view
this coincidence as a  confirmation of the choices \ALEGep\ and \SWGep\
for the Landau-Ginzburg potentials, and as a further successful test on how
BCFT techniques work in non-compact situations.

Rather, the important point is that the spectrum of the $D$-brane
boundary states turns out to be truncated exactly such that it matches
the expected dyon spectrum of the strongly coupled Yang-Mills theory,
and this is more than just simple algebraic geometry and group theory.
In order to appreciate this, recall that the geometry of a compact or
non-compact CY manifold determines a priori only the homology lattice
and thus what the {\it possible} RR charges  of wrapped branes are.
However, it does not directly tell what subset of the charge lattice
corresponds to the stable, single
particle quantum states, at a given point in the moduli
space.\foot{For recent considerations about this issue, see \DFR.}

As has been demonstrated in the present and in other recent papers, 
this kind of questions can be
analyzed in terms of boundary conformal field theory,  
at least as far as exactly solvable
models are concerned. In particular we have found that
the BPS spectrum of stable $D$-branes on ALE spaces and SW curves,
at the Gepner points of their respective moduli spaces,
can be mapped one-to-one  
to the spectrum of primary fields of the
$\N=2$ superconformal minimal models. The latter is finite
due to the truncation to $\ell\leq k$ of the fusion rules,
which manifests itself in the sine functions in the fusion
coefficients~\FR; this is intimately related to the truncation
to integrable representations of $\widehat{SU}(2)_k$. 

The fact that these two instances of quantum truncation, namely of the
spectrum of wrapped $D$-branes on the one hand and of the spectrum of
2d primary fields on the other, can be directly mapped into each other,
is what we view as the most interesting aspect of our
considerations.

There are certain obvious generalizations one
could think about, like considering $\N=2$ $d=4$ gauge theories
with different gauge groups and matter content. 
As for pure Yang-Mills theories based on simply laced Lie algebras
of type $ADE$, we expect the following Landau-Ginzburg
potentials to provide a useful BCFT formulation:
$$
W^{SW}_{ADE}(z_i,x_j,u_k)\ =\ 
{1\over {z_1}^{2h(ADE)}}+ {1\over {z_2}^{2h(ADE)}}+P_{ADE}(x_j,u_k)\ .
\eqn\SWGep
$$
Here, $h(ADE)$ denotes the dual Coxeter number and $P_{ADE}$ the normal
form of the simple singularity of the corresponding type \arn. 

Moreover it may be interesting to find a relation between our and the
work of \DHT, where a correspondence between BPS  spectra of certain
two and four dimensional $\N=2$ supersymmetric gauge theories has been
discovered.   Finally one may ask about the significance of integrable
deformations \integ\  of the $\N=2$ minimal models in the present
context; these leave infinitely many two-dimensional currents
conserved, and so one may wonder about enhanced integrability
properties of the Yang-Mills theories on certain sub-loci on the moduli
space.

\goodbreak 
%%%%%%%%%%%%%%%%%%%%%%%%%%%%%%%%%%%%%%%%%%%%%%%%%%%%%%%
\ack
I like to thank Peter Kaste, Andy L\"utken and Johannes Walcher
for discussions on various
aspects of boundary conformal field theory.

%%%%%%%%%%%%%%%%%%%%%%%%%%%%%%%%%%%%%%%%%%%%%%%%%%%%%%%
\nobreak

\bigskip
\goodbreak
\refout
\end